\documentclass[useAMS,usenatbib,usegraphicx]{mn2e}

\newcommand{\pdif}[2]{\frac{\rmn{\partial} #1}{\rmn{\partial} #2}}

\title[Mass distribution of NGC~4594]{Line-of-sight velocity dispersions and
a mass distribution model of the Sa galaxy NGC~4594}
\author[E. Tempel and P. Tenjes]{E. Tempel$^{1,2}$\thanks{E-mail: elmo@aai.ee;
peeter.tenjes@ut.ee} and P. Tenjes$^{1,2}$\\
$^{1}$ Institute of Theoretical Physics, Tartu University, T\"ahe 4,
51010 Tartu, Estonia\\
$^{2}$ Tartu Astrophysical Observatory, 61602 T\~oravere, Estonia}

\begin{document}

\date{Accepted 2006 June 27. Received 2006 June 21; in original form 2006 April 25.}

\pagerange{\pageref{firstpage}--\pageref{lastpage}} \pubyear{2006}

\maketitle

\label{firstpage}

\begin{abstract}
In the present paper we develop an algorithm allowing to calculate
line-of-sight velocity dispersions in an axisymmetric galaxy outside
of the galactic plane. When constructing a self-consistent model, we
take into account the galactic surface brightness distribution,
stellar rotation curve and velocity dispersions. We assume that the
velocity dispersion ellipsoid is triaxial and lies under a certain
angle with respect to the galactic plane. This algorithm is applied
to a Sa galaxy NGC~4594~= M~104, for which there exist velocity
dispersion measurements outside of the galactic major axis. The mass
distribution model is constructed in two stages. In the first stage
we construct a luminosity distribution model, where only galactic
surface brightness distribution is taken into account. Here we
assume the galaxy to consist of the nucleus, the bulge, the disc and
the stellar metal-poor halo and determine structure parameters of
these components. Thereafter, in the second stage we develop on the
basis of the Jeans equations a detailed mass distribution model and
calculate line-of-sight velocity dispersions and the stellar
rotation curve. Here a dark matter halo is added to visible
components. Calculated dispersions are compared with observations
along different slit positions perpendicular and parallel to the
galactic major axis. In the best-fitting model velocity dispersion
ellipsoids are radially elongated with $\sigma_{\theta}/\sigma_R
\simeq 0.9-0.4$, $\sigma_z/\sigma_R \simeq 0.7-0.4$, and lie under
the angles $\le 30\degr$ with respect to the galactic equatorial
plane. Outside the galactic plane velocity dispersion behaviour is
more sensitive to the dark matter density distribution and allows to
estimate dark halo parameters. For visible matter the total $M/L_B =
4.5 \pm 1.2$, $M/L_R = 3.1 \pm 0.7$. The central density of the dark
matter halo is $\rho_{\rm DM}(0) = 0.033~\rmn{M_{\sun}pc^{-3}}$.
\end{abstract}


\section{Introduction}

The study of the dark matter (DM) halo density distribution allows
us to constrain possible galaxy formation models and large scale
structure formation scenarios \citep*{b66,b51,b41}. For this kind of
analysis, it is necessary to know both the distribution of visible
and dark matter. Without additional assumptions rotation curve data
alone are not sufficient to discriminate between these two kinds of
matter \citep{b29}. It does not suffice either to use additionally
velocity dispersions along the major axis.

Realistic mass and light distribution models must be consistent,
i.e. the same model must describe the luminosity distribution and
kinematics. Three main classes of self-consistent mass distribution
models can be discriminated: the Jeans equations based models, the
specific phase space density distribution models and the
Schwarzschild orbit superposition based models.

Mass distribution models based on solving the Jeans equations have
an advantage that the equations contain explicitly observed
functions -- velocity dispersions. On the other hand, there are
three equations, but at least five unknown functions (three
dispersion components, centroid velocity and the velocity dispersion
ellipsoid orientation parameter) and thus the system of equations is
not closed. In addition, the use of the Jeans equations neglects
possible deviations of velocity distributions from Gaussians and
does not garantee that the derived dynamical model has non-negative
phase density distribution everywhere. However, within certain
approximations the Jeans equations are widely used for the
construction of mass distribution models. In the case of spherical
systems with biaxial velocity dispersion ellipsoids, such models
have been constructed, for example, by \citet{b6}, \citet{b64},
\citet{b42}, \citet{b82}. In the case of flattened systems with
biaxial velocity dispersion ellipsoids, a general algorithm for the
solution of the Jeans equations was developed by \citet*{b8},
\citet{b19}. Another algorithm in the context of the multi-Gaussian
expansion (MGE) formalism was developed by \citet{b34}. An
approximation for cool stellar discs (random motions are small when
compared with rotation) has been developed by \citet{b1}.

Dynamical models with a specific phase density distribution have the
advantage that velocity dispersion anisotropy can be calculated
directly. On the other hand, due to rather complicated analytical
calculations, only rather limited classes of distribution functions
can be studied. Spherical models of this kind have been constructed
by \citet*{b17}, \citet{b5}. In the case of an axisymmetric density
distribution, velocity dispersion profiles have been calculated for
certain specific mass and phase density distribution forms by
\citet*{b84}, \citet{b37}, \citet{b25}, \citet*{b24}, \citet*{b28},
\citet{b65}, \citet{b2} and others. A special case is an analytical
solution with three integrals of motion for some specific
potentials: an axisymmetric model with a potential in the St\"ackel
form \citep{b27}, isochrone potential \citep{b26}.

Probably the most complete class of dynamical models has been
developed on the basis of the Schwarzschild linear programming
method \citep{b72}. Thus, it is not surprising that just for this
method most significant developments occured in last decade.
\citet{b68} and \citet{b22} have developed this method in order to
calculate line-of-sight velocity profiles. Thereafter, \citet{b16}
and \citet{b88} generalized it for an arbitrary density distribution
linking it with MGE method. A modification of the least-square
algorithm was done by \citet{b55}. Interesting comparisons of the
results of the Schwarzschild method with phase density calculations
within a two-integral approximation have been made by \citet{b86}
and \citet{b55}.

At present, nearly all dynamical models have been applied for
one-component systems. However, the structure of real galaxies is
rather complicated -- galaxies consist of several stellar
populations with different density distribution and different
ellipticities. In addition, in different components velocity
dispersions or rotation may dominate.

In our earlier multi-componental models \citep*[see][]{b79,b80,b32}
we approximated flat components with pure rotation models and
spheroidal components with dispersion dominating kinematics. For
spheroidal components mean velocity dispersions were calculated only
on the basis of virial theorem for multi-componental systems. These
models fit central velocity dispersions, gas rotation velocities and
light distribution with self-consistent models.

In the present paper, we construct a more sophisticated
self-consistent mass and light distribution model. We decided to
base it on the Jeans equations. For all visible components, both
rotation and velocity dispersions are taken into account. The
velocity dispersion ellipsoid is assumed to be triaxial and
line-of-sight velocity dispersions are calculated. Mass distribution
of a galaxy is axisymmetric and inclination of the galactic plane
with respect to the plane of the sky is arbitrary.

In order to discriminate between DM and visible matter, it is most
complicated to determine the contribution of the stellar disc to the
galactic mass distribution. Quite often the maximum disc
approximation is used. In the present paper, we attempt to decrease
degeneracy, comparing calculated models with the observed stellar
rotation curve, velocity dispersions along the major axis and in
addition, along several cuts parallel to the major and minor axis.
In the case of two-integral models for edge-on galaxies this allowed
to constrain possible dynamical models \citep{b63}.

First measurements of velocity dispersions along several slit
positions were made by \citet{b53} and \citet{b46}. Later, similar
measurements were performed by \citet{b8}, \citet*{b38},
\citet{b77}, \citet{b16}. In recent years, with the help of integral
field spectroscopy, complete 3D velocity and dispersion fields have
been measured already for several tens of galaxies.

We apply our model for the nearby spiral Sa galaxy M~104 (NGC~4594,
the Sombrero galaxy). This galaxy is suitable for model testing,
being a disc galaxy with a significant spheroidal component. The
galaxy has a detailed surface brightness distribution and a
well-determined stellar rotation curve. M~104 has a significant
globular cluster (GC) subsystem. And as most important in our case,
the line-of-sight velocity dispersion has been measured along the
slit at different positions parallel and perpendicular to the
projected major axis.

We construct the model in two stages. First, a surface brightness
distribution model is calculated. Here we distinguish stellar
populations and calculate their structure parameters with the
exception of masses. In the second stage, we calculate line-of-sight
velocity dispersions and the stellar rotational curve and derive a
mass distribution model.

Sections~2 and 3 describe the observational data used in the
modelling process and construction of the preliminary model. In
Section~4 we present the line-of-sight dispersion modelling process.
Section~5 is devoted to the final M~104 modelling process. In
Section~6 the discussion of the model is presented.

Throughout this paper all luminosities and colour indices have been
corrected for absorption in our Galaxy according to \citet*{b71}.
The distance to M~104 has been taken 9.1~Mpc, corresponding to the
scale 1~arcsec~= 0.044~kpc \citep*{b39,b61,b81}. The angle of
inclination has been taken 84\degr.

\section{Observational data used}
By now a surface photometry of M~104 is available in $UBVRIJHK$
colours. In the present study, we do not use the $U$-profile, as
this profile has a rather limited spatial extent and is probably
most significantly distorted by absorption. In certain regions also
the $B$-profile is probably influenced by absorption, but the
$B$-profile has the largest spatial extent and we decided to use it
with some caution outside prominent dust lane absorption. The
$JHK$-profiles have a rather limited spatial extent and resolution
and we decided not to use them in here. In this way the surface
brightness profiles in $BVRI$ colours were compiled. Different
colour profiles help to distinguish stellar populations and allow to
calculate corresponding $M/L$ ratios, and thereafter, colour indices
of the components. Table~\ref{phot_dat} presents references, the
faintest observed isophotes ($\rmn{mag~arcsec^{-2}}$), and the
colour system used. Different $R$ colour system profiles are
transferred into the Cousins system, using the calibration by
\citet{b40}. The observations by \citet{b76} were made without
absolute calibration. They were calibrated with the help of other
$R$ colour observations. \citet{b45} observed M~104 in $BVRI$
colours but in their paper only colour indices are given and we
cannot use them in here.

\begin{table}
\caption{Photometrical data} \label{phot_dat}
\begin{flushleft}
\begin{tabular}{llll}
\hline\noalign{\smallskip}
Reference                           & Faintest & Colour \\
                                    & isophote & system \\
\noalign{\smallskip} \hline\noalign{\smallskip}
\citet{b76}            & ~        & $R$ \\
\citet{b14}            & 28.2     & $B$ \\
\citet{b10}            & 24.7     & $B$ \\
\citet{b44}            & 22.7     & $B$ \\
\citet{b4}             & 27.3     & $B$ \\
\citet{b47}            & 25.0     & $V$ \\
\citet{b15}            & 23.1     & $BVRI$ \\
\citet{b49}            & 23.3     & $R$ \\
\citet{b52}            & 19.2     & $V$ \\
\citet{b21}            & 17.4     & $B$ \\
\citet{b34}            & 18.5     & $R$ \\
\citet{b35}            & 17.8     & $V$ \\
\noalign{\smallskip} \hline
\end{tabular}
\end{flushleft}
\end{table}

\begin{figure}
\includegraphics[width=84mm]{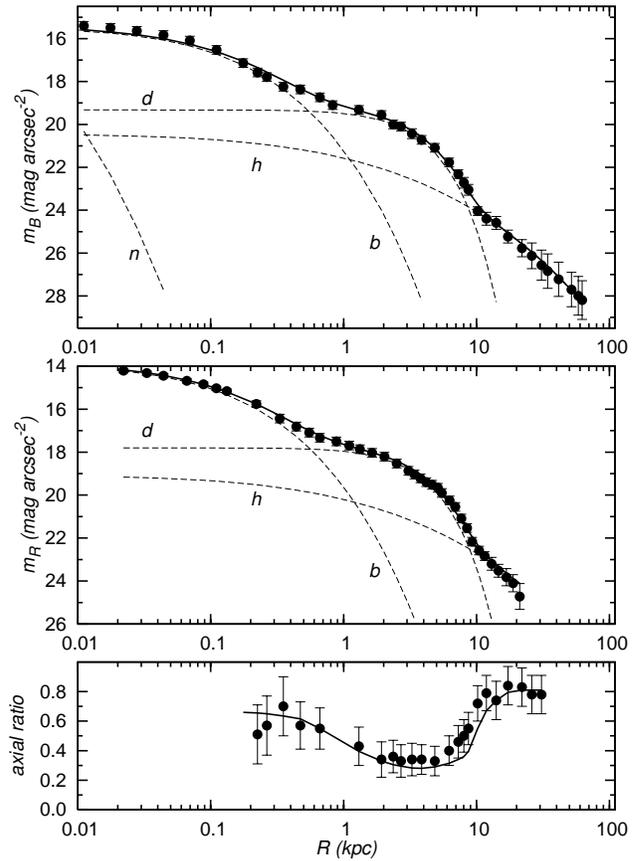} \caption{Upper
panels: the averaged surface brightness profiles of M~104 in the $B$
and $R$ colours. Filled circles -- observations, solid line --
model, dashed lines -- models for components (n -- the nucleus, b --
the bulge, h -- the metal-poor halo, d -- the disc). Lower panel:
the axial ratios of M~104 isophotes as a function of the
galactocentric distance.} \label{photom}
\end{figure}

The composite surface brightness profiles in the $BVRI$ colours
along the major and/or the minor axes were derived by averaging the
results of different authors. Due to dust absorption lane, surface
brightnesses only on one side along the minor axis have been taken
into account. All the surface brightness profiles obtained in this
way belong to the initial data set of our model construction. To
spare space, we present here the surface brightness distributions in
$B$ and $R$ only (Fig.~\ref{photom} upper panels), and the axial
ratios (the ratio of the minor axis to the major axis of an
isophote) (Fig.~\ref{photom} lower panel) as functions of the
galactocentric distance.

\begin{figure}
\includegraphics[width=84mm]{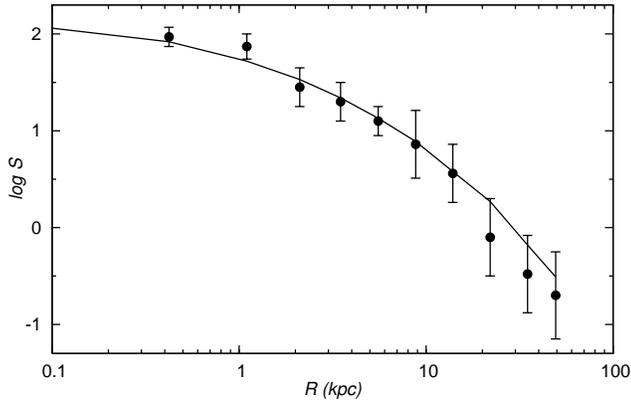} \caption{The
surface density distribution of GCs in M~104. The observations are
presented by filled circles. The continuous line gives our
best-fitting model distribution for the halo.} \label{glob_cl}
\end{figure}

The observed surface density distribution of GC candidates was
derived by \citet{b11}, \citet{b61} and \citet{b67}. The derived
distributions were averaged, taking into account different
background levels. The resulting surface density distribution of GC
candidates is given in Fig.~\ref{glob_cl} by filled circles and was
used to constrain stellar metal-poor halo parameters. Line-of-sight
velocities of GCs were measured by \citet{b12} and the calculated
mean velocity dispersion of GC subsystem $\sigma_{\rmn{GC}}=
\rmn{255~km~s^{-1}}$ was derived.

\begin{figure}
\includegraphics[width=84mm]{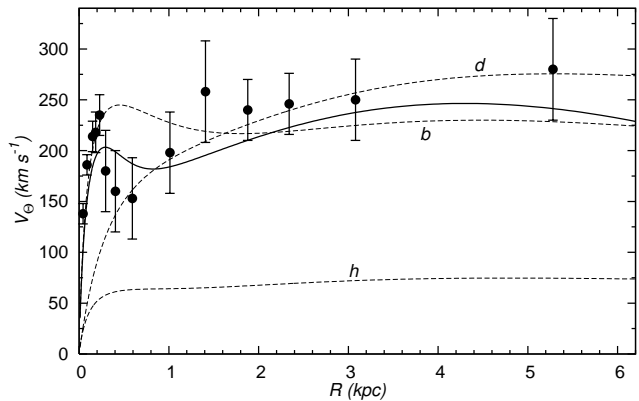} \caption{The
rotation curve of M~104 on the basis of stellar rotation velocities.
Filled circles -- observations, solid line -- model, dashed lines --
models for components.} \label{stelrot}
\end{figure}

\begin{figure}
\includegraphics[width=84mm]{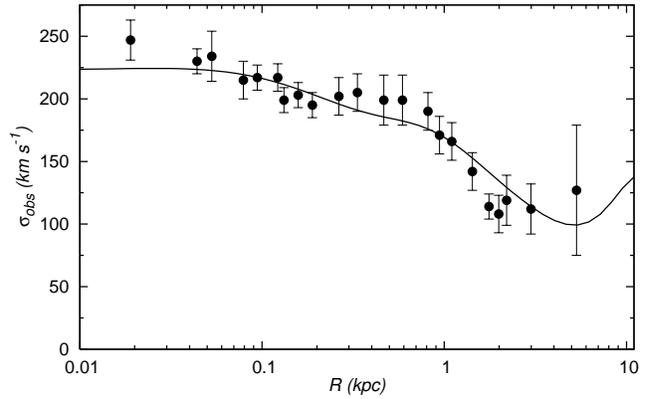}
\caption{Observed line-of-sight velocity dispersions of M~104 along
the major axis. Filled circles -- observations, solid line --
model.} \label{dispmaj}
\end{figure}

Rotation velocities of stars and line-of-sight velocity dispersion
profile along the major axis in very good seeing conditions
(0.2--0.4~arcsec) for the central regions was obtained by
\citet{b54} with {\it HST\/} and {\it CFHT\/}. In addition, the
central regions were measured by \citet{b18}, \citet{b35}. In the
central and intermediate distance interval, dispersions and stellar
rotation have been measured by \citet{b53}, \citet{b45} and
\citet{b85}. We averaged the stellar rotation velocities at various
distance intervals with weights depending on seeing conditions and
velocity resolution, and derived the stellar rotation curve
presented by filled circles in Fig.~\ref{stelrot}. Averaged in the
same way line-of-sight velocity dispersions along the major axis are
presented by filled circles in Fig.~\ref{dispmaj}.

In addition, \citet{b53} derived dispersion profiles along several
slit positions (at  0, 30, 40, 50~arcsec parallel and at 0,
50~arcsec perpendicular to the major axis) in the bulge component.
We use them in mass distribution model calculations (filled circles
in Figs.~\ref{dispmajor},~\ref{dispminor}).

\begin{figure}
\includegraphics[width=84mm]{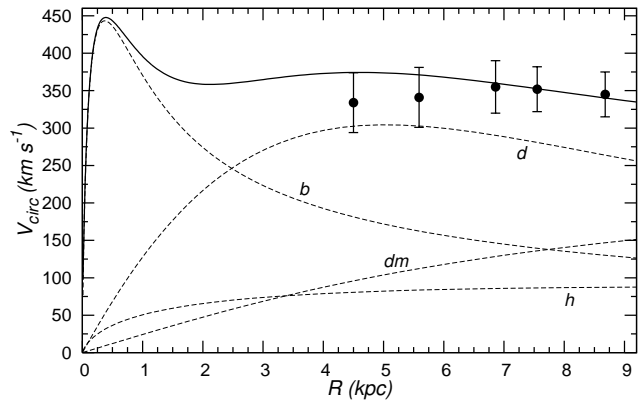}
\caption{Calculated circular velocity for the best-fitting model of
M~104 (solid line). Dashed lines give circular velocities for
components (dm -- dark matter). Observed gas velocities are given by
filled circles. Only outermost points are given where stellar
motions are not known.} \label{gasrot}
\end{figure}

Ionized gas radial velocities were obtained and the rotation curve
was constructed by \citet{b73} and \citet{b69}. HI velocities at
11.4~arcsec (0.5~kpc) resolution were obtained by \citet{b3}.
Unfortunately, we have no detailed information about the gas
velocity dispersions. When using gas rotation velocities, often an
assumption is made that gas dispersions are small when compared with
rotation velocities, and in this way, rotation velocities are taken
to be circular velocities. However, in the case of M~104, up to
distances ~$\sim 3$~kpc, rotation velocities of stars and gas are
comparable and thus we may expect also dispersions to be comparable,
and therefore, gas dispersions cannot be neglected. For this reason,
we cannot use gas rotation velocities directly in fitting the model.
We use gas rotation only to have an approximate mass distribution
estimate at large galactocentric distances where stellar rotation
and dispersion data do not extend. In these outer parts, the
velocities from different studies were averaged and the resulting
gas rotation velocities are given by filled circles in
Fig.~\ref{gasrot}.
\section{Surface brightness distribution model of the M~104}
To construct a model of the M~104 galaxy, we limit the main stellar
components to the central nucleus, the bulge, the disc and the
metal-poor halo. To construct  a dynamical model in the following
sections, a DM component --~the dark halo~-- must be added to
visible components.

To construct the light distribution model, the surface luminosity
distribution of components is usually approximated by the S\'{e}rsic
formula \citep{b74}. If, in addition to the photometrical data,
kinematic data are also used, the corresponding dynamical model must
be consistent with the photometry, i.e. the same density
distribution law must be used for rotation curve modelling (and for
the velocity dispersion curve, if possible). For spherical systems,
an expression for circular velocity with an integer S\'{e}rsic index
can be derived \citep{b62}. For a non-integer index and ellipsoidal
surface density distribution, a consistent solution for rotation
curve calculations is not known.

In the present paper, the density distribution parameters are
determined by the least squares method and may have any value. In
addition, our intention is to use the model also for velocity
dispersion calculations.

For the reasons given above, we decided to construct models starting
from a spatial density distribution law for individual components,
which allows an easier fitting simultaneously for light distribution
and kinematics.

In such models, the visible part of a galaxy is given as a
superposition of the nucleus, the bulge, the disc and the metal-poor
halo. The spatial density distribution of each visible component is
approximated by an inhomogeneous ellipsoid of rotational symmetry
with the constant axial ratio $q$ and the density distribution law
\begin{equation}
l (a)=l (0)\exp [ -( a/(ka_0))^{1/N} ] , \label{eq1}
\end{equation}
where $l (0)=hL/(4\pi q a_0^3)$ is the central density and $L$ is
the component luminosity; $a= \sqrt{R^2+z^2/q^2}$, where $R$ and $z$
are two cylindrical coordinates, $a_0$ is the harmonic mean radius
which characterizes rather well the real extent of the component,
independently of the parameter $N$. Coefficients $h$ and $k$ are
normalizing parameters, depending on $N$, which allows the density
behavior to vary with $a$. The definition of the normalizing
parameters $h$ and $k$ and their calculation is described in
appendix B of \citet{b79}. Equation~(\ref{eq1}) allows a
sufficiently precise numerical integration and has a minimum number
of free parameters.

The density distributions for the visible components were projected
along the line-of-sight, and their sum gives us the surface
brightness distribution of the model
\begin{equation}
L(A) = 2 \sum_{i=1}^4 {q_i\over Q_i} \int\limits_A^{\infty}
          {l_i (a) a~\rmn{d}a\over (a^2 - A^2)^{1/2}} ,
\label{eq2}
\end{equation}
where $A$ is the major semiaxis of the equidensity ellipse of the
projected light distribution and $Q_i$ are their apparent axial
ratios $Q^2=\cos^2\delta +q^2\sin^2\delta$. The angle between the
plane of the galaxy and the plane of the sky is denoted by $\delta$.
The summation index $i$ designates four visible components.

For the nucleus and the stellar metal-poor halo, parameters $q$,
$a_0$, $N$ were determined independently of other subsystems. For
the nucleus, these parameters were determined on the basis of the
central light distribution; for the metal-poor halo, these
parameters were determined on the basis of the GC distribution. In
subsequent fitting processes, these parameters were kept fixed. This
step allows to reduce the number of free parameters in the
approximation process.

The model parameters $q$, $a_0$, $L$, and $N$ for the bulge and the
disc, and $L$ for the nucleus and the halo were determined by a
subsequent least-squares approximation process. First, we made a
crude estimation of the population parameters. The purpose of this
step is to avoid obviously non-physical parameters -- relation
(\ref{eq2}) is non-linear and fitting of the model to observations
is not a straightforward procedure. Next, a mathematically correct
solution was found. Details of the least squares approximation and
the general modelling procedure were described by \citet{b31},
\citet{b79,b80}.

Total number of free parameters (degrees of freedom) in
least-squares approximation was 18, the number of observational
points was 231. Transition from the bulge to the disc and from the
disc to the metal-poor halo is rather well-determined by comparing
the light profiles along the major and the minor axis (see
Fig.~\ref{photom} lower panel). Parameters of the nucleus are more
uncertain because no sufficiently high-resolution central luminosity
distribution observations are available for us. On the other hand
our aim is to study general mass distribution in M~104 where nuclear
contribution in small. For this reason convolution and deconvolution
processes were not used in luminosity distribution model and in
subsequent mass distribution model.

The final parameters of the model (the axial ratio $q$, the harmonic
mean radius $a_0$,  the structural parameters $N$, the dimensionless
normalizing constants $h$ and $k$, $BVRI$-luminosities) are given in
Table~\ref{model_param}. The model is represented by solid lines in
Figs.~\ref{photom},~\ref{glob_cl}. The mean deviation of the model
from the observations of surface brightnesses is $\langle
\mu^\rmn{obs} -\mu^\rmn{model} \rangle = 0.16~\rmn{mag}$.

\section{Calculation of velocity dispersions}
\subsection{Basic formulae}
Knowing spatial luminosity densities of the components $l_i(a)$ and
ascribing a mass-to-light ratios to each component $f_i$ ($i$
indexes the nucleus, the bulge, the disc and the stellar metal-poor
halo), we have spatial mass density distribution of a galaxy
\begin{equation}
\rho (a) = \sum_{i=1}^4 f_i l_i(a) + \rho_{\rm DM}(a) \label{eq3}
\end{equation}
($\rho_{\rmn{DM}} (a)$ is the DM density). On the basis of spatial
mass density distributions, derivatives of the gravitational
potential $\pdif{\Phi}{R}$ and $\pdif{\Phi}{z}$ can be calculated
\citep[see][]{b7}.

In stationary collisionless stellar systems with axial symmetry the
Jeans equations in cylindrical coordinates are
\begin{equation}
\pdif{(\rho\overline{v^2_R})}{R} + \pdif{(\rho\overline{v_R
v_z})}{z} + \rho\left(\frac{\overline{v^2_R} -
\overline{v^2_\theta}}{R} + \pdif{\Phi}{R}\right)=0,
\end{equation}
\begin{equation}
\pdif{(\rho\overline{v_R v_\theta})}{R} +
\pdif{(\rho\overline{v_\theta v_z})}{z} +
\frac{2\rho}{R}\overline{v_\theta v_R}=0, \label{j2}
\end{equation}
\begin{equation}
\pdif{(\rho\overline{v_R v_z})}{R} +
\pdif{(\rho\overline{v^2_z})}{z} + \frac{\rho\overline{v_R v_z}}{R}
+ \rho\pdif{\Phi}{z}=0,
\end{equation}
where $v_R$, $v_z$, $v_\theta$ are velocity components.

\begin{figure}
\includegraphics[width=84mm]{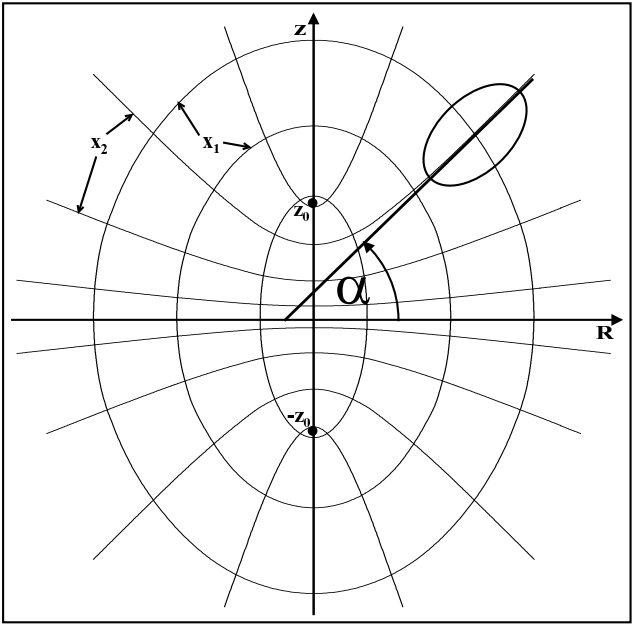}
\caption{Elliptical coordinates ($x_1$, $x_2$) and their relations
with cylindrical coordinates ($R$, $z$) in galactic meridional
plane.}\label{ell_coord}
\end{figure}

The velocity dispersion tensor $\sigma^2_{ij} = \overline{v_i
v_j}-\overline{v}_i\overline{v}_j$ in the diagonal form for the
axisymmetric case can be described by four variables: dispersions
along the coordinate axis ($\sigma_R$, $\sigma_z$ and
$\sigma_\theta$) and an orientation angle $\alpha$ in $Rz$-plane
(see Fig.~\ref{ell_coord}). Mixed components of the tensor are
\begin{equation}
\sigma^2_{Rz}=\gamma(\sigma^2_R-\sigma^2_z), \qquad
\sigma^2_{R\theta}=\sigma^2_{z\theta}=0,
\end{equation}
where
\begin{equation}
\gamma=\frac{1}{2}\tan{2\alpha}.
\end{equation}
As a result of axial symmetry the second Jeans equation (\ref{j2})
vanishes.

Introducing the dispersion ratios
\begin{equation}
k_z\equiv\frac{\sigma^2_z}{\sigma^2_R}, \qquad
k_\theta\equiv\frac{\sigma^2_\theta}{\sigma^2_R},
\end{equation}
the remaining Jeans equations can be written in a more convenient
form for us
\begin{eqnarray}
\pdif{\rho\sigma^2_R}{R} + \left( \frac{1-k_\theta}{R} +
\pdif{\kappa}{z} \right)\rho\sigma^2_R +
\kappa\pdif{\rho\sigma^2_R}{z} =\nonumber\\
= -\rho\left(\pdif{\Phi}{R} - \frac{V^2_\theta}{R} \right),
\label{jj1}
\end{eqnarray}
\begin{equation}
\pdif{\rho\sigma^2_z}{z} + \left( \frac{\xi}{R} + \pdif{\xi}{R}
\right)\rho\sigma^2_z + \xi\pdif{\rho\sigma^2_z}{R} =
-\rho\pdif{\Phi}{z}, \label{jj2}
\end{equation}
where $V_{\theta}$ is rotational velocity,
\begin{equation}
\kappa\equiv\gamma(1-k_z), \qquad \xi\equiv\frac{\kappa}{k_z}.
\end{equation}
For each component the rotation velocity have been taken $V_{\theta}
= \beta V_{c}$, where $V_c$ is circular velocity and $\beta$ is a
constant specific for each subsystems. Taking into account the
definition of the circular velocity we can substitute in
Eq.~(\ref{jj1})
\begin{equation}
\frac{V_\theta^2}{R}=\beta^2 \pdif{\Phi}{R}.
\end{equation}

The Jeans equations (\ref{jj1}) and (\ref{jj2}) include unknown
values $k_z$, $k_\theta$ and $\gamma$. Spatial density,
gravitational potential and rotational velocity can be determined on
the basis of the galactic surface brightness distribution
(Eq.~\ref{eq2}), the Poisson equation and the observed stellar
rotation curve. Dispersions $\sigma^2_R$ and $\sigma^2_z$ must be
calculated from the Jeans equations.

Following the notation of \citet{b60}, we define confocal elliptic
coordinates ($x_1$, $x_2$) as the roots of
\begin{equation}
\frac{R^2}{x^2-z^2_0} + \frac{z^2}{x^2}=1,
\end{equation}
where
\begin{equation}
x^2=\left\{\begin{array}{cc} x_1^2 & \geq  z_0^2 \\ x_2^2 & \leq
z_0^2 \end{array}  \right  . .
\end{equation}
The third coordinate $x_3=\theta$. Foci of ellipses and hyperbolae
are determined by $\pm z_0$. The relations between elliptic and
cylindrical coordinates are as follows:
\begin{equation}
x^2_1=\frac{1}{2}\left[\Omega + (\Omega^2 -
4z^2z^2_0)^{1/2}\right],
\end{equation}
\begin{equation}
x^2_2=\frac{1}{2}\left[\Omega - (\Omega^2 -
4z^2z^2_0)^{1/2}\right],
\end{equation}
where
\begin{equation}
\Omega\equiv R^2 + z^2_0 + z^2.
\end{equation}
Later we also need the relation
\begin{equation}
\tan\alpha = \frac{z^2 - x^2_2}{Rz}.
\end{equation}

In this case, the parameter $\gamma$ related to the angle between
the ellipsoid major axis and the galactic disc is
\begin{equation}
\gamma = \frac{Rz}{R^2+z^2_0-z^2}.
\end{equation}
The position of foci $z_0$ is at present a free parameter, which
must be determined within the modelling process. For $z_0= const$
the orientation of the velocity ellipsoid would be along the
elliptic coordinates. Velocity dispersions along elliptical
coordinates ($x_1$, $x_2$, $x_3$) are denoted as ($\sigma_1$,
$\sigma_2$, $\sigma_3$).

In the case of a triaxial velocity ellipsoid, the phase density of a
stellar system is a function of three integrals of motion. For an
axisymmetrical system, in addition to energy and angular momentum
integrals, a third non-classical integral is needed. As it was
stressed by \citet{b57}, this third integral should be quadratic
with respect to velocities (in this case minimum number of
constraints result for gravitational potential). On the basis of
this assumption, \citet{b57} derived a corresponding form of the
third integral.

Starting from the form of Kuzmin's third integral, \citet{b30}
derived that dispersion ratios can be written in the form
\begin{equation}
k_{12} \equiv\frac{\sigma^2_2}{\sigma^2_1} = \frac{a_1z^2_0 +
a_2x^2_2}{a_1z^2_0+a_2x^2_1},
\end{equation}
\begin{equation}
k_{13}\equiv\frac{\sigma^2_3}{\sigma^2_1} = \frac{a_1z^2_0 +
a_2x^2_2}{a_1z^2_0 + a_2z^2 + b_2R^2},
\end{equation}
where $a_1$, $a_2$ and $b_2$ are unknown parameters. As a
simplifying assumption these three parameters were taken in
\citet{b30} $a_1=a_2=b_2$. In the present paper we determine these
parameters by demanding that $a_1$, $a_2$ and $b_2$ must satisfy the
relation \citep{b58}
\begin{equation}
\frac{1}{k_{12}}=1+\frac{1}{k_{13}}. \label{kuz1}
\end{equation}
This relation was derived by Kuzmin in the case of disclike systems
and we must keep in mind that therefore our results may not be a
good approximation far from the galactic plane.

Using the relations between cylindric and elliptic coordinates we
derive
\begin{equation}
k_z = \frac{\sin^2\alpha + k_{12}\cos^2\alpha}{\cos^2\alpha +
k_{12}\sin^2\alpha} = \frac{\tan^2\alpha +
k_{12}}{1+k_{12}\tan^2\alpha},
\end{equation}
\begin{equation}
k_\theta = \frac{k_{13}}{\cos^2\alpha + k_{12}\sin^2\alpha} =
\frac{k_{13}(1+\tan^2\alpha)}{1+k_{12}\tan^2\alpha}.
\end{equation}

The quantity $z_0$ determines the orientation of the velocity
ellipsoid. For specific density distribution (gravitational
potential) forms within the theory of the third integral of motion
$z_0=const$. In the case of general density distributions
$z_0=f(R,z)$. For example, it was demonstrated by \citet{b59} that
in a galactic disc
\begin{equation}
R\left(\pdif{\gamma}{z} \right)_{z=0} =
-\frac{1}{4}\frac{\rmn{\partial} \ln \rho_t}{\rmn{\partial}\ln R},
\label{kuz2}
\end{equation}
where $\rho_t$ is total galactic spatial mass density. From
Eq.~(\ref{kuz2}) we can determine $z_0^2$
\begin{equation}
z^2_0(R,0)=-R\left[4\rho_t(R,0)\left(\pdif{\rho_t(R,0)}{R
}\right)^{-1}+R\right].
\end{equation}
The dependence of $z_0$ on $z$ is derived to have the best-fitting
with measured dispersions.
\subsection{Line-of-sight dispersions}
\begin{figure}
\includegraphics[width=84mm]{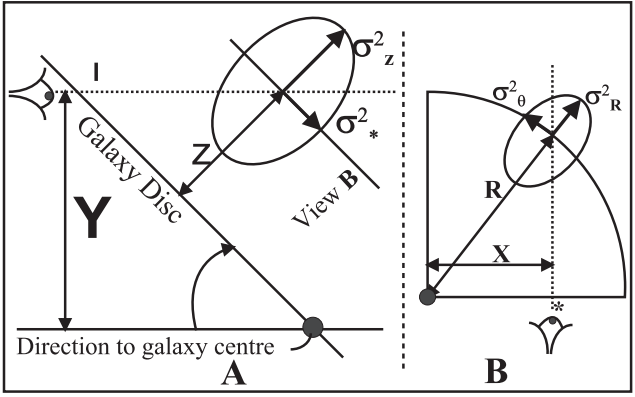}
\caption{Dispersion projection in a plane parallel to galactic
disc.}\label{proj1}
\end{figure}

\begin{figure}
\includegraphics[width=84mm]{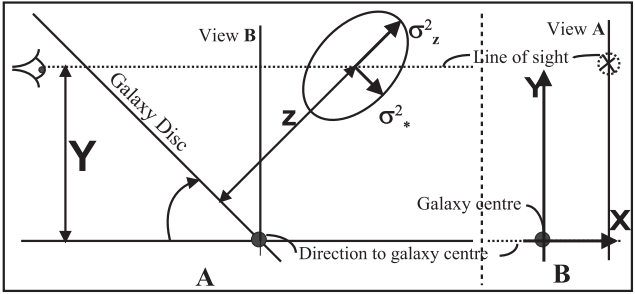}
\caption{Projection of dispersions to the
line-of-sight.}\label{proj2}
\end{figure}

Calculated on the basis of hydrodynamic models, dispersions
$\sigma^2_R$, $\sigma^2_z$ and $\sigma^2_\theta$ cannot be compared
directly with measurements. We project the velocity dispersions in
two steps (see Figs.~\ref{proj1},~\ref{proj2}). First, we make a
projection in a plane parallel to the galactic disc. For this we
must project $\sigma^2_R$ and $\sigma^2_\theta$ to the disc, going
along the line-of-sight and being parallel to the galactic disc.
Projected dispersions are
\begin{equation}
\sigma^2_\ast=\sigma^2_\theta\frac{X^2}{R^2} + \sigma^2_R \left(
1-\frac{X^2}{R^2}\right) .
\end{equation}
Second, we must project dispersions $\sigma^2_z$ and $\sigma^2_\ast$
to the line-of-sight. Designating $\Theta$ as the angle between the
line-of-sight and the galactic disc, the line-of-sight dispersion
$\sigma^2_l$ is
\begin{equation}
\sigma^2_l=\sigma^2_\ast\cos^2\Theta + \sigma^2_z\sin^2\Theta .
\end{equation}

To compare the calculated dispersions with the measured data, we
must calculate averaged along the line-of-sight dispersion.
Integrating dispersions along the line-of-sight we may write
\begin{equation}
\sigma^2_{int}(X,Y)=\frac{1}{L(X,Y)}
\int\limits^{\infty}_{-\infty}l(R,z)\sigma^2_l(R,z)\rmn{d}l,
\end{equation}
where $l(R,z)$ denotes galactic spatial luminosity density, and
$L(X,Y)$ is the surface luminosity density profile (please note that
integration $\rmn{d}l$ means integration along the line-of-sight).
Changing variables in the integral to have integration along the
radius, we obtain
\begin{equation} \label{final_eq}
\sigma^2_{int}(X,Y)=\frac{1}{L(X,Y)}\int\limits^{\infty}_{X} \Psi
\frac{R}{\cos{\Theta} \sqrt{R^2-X^2}}\rmn{d}R,
\end{equation}
where
\begin{equation}
\Psi\equiv l(R,z_1)\sigma^2_l(R,z_1)+ l(R,z_2)\sigma^2_l(R,z_2),
\end{equation}
\begin{equation}
z_{1,2}= \left( \frac{Y}{\sin{\Theta}} \pm \sqrt{R^2-X^2} \right)
\tan{\Theta}.
\end{equation}
Equation~(\ref{final_eq}) gives the line-of-sight dispersion for one
galactic component. As in our model we have several components, we
must sum over all components considering the luminosity distribution
profile
\begin{equation}
\sigma_{\rm obs}(X,Y)=\left\{\frac{\sum\limits_i\left[L_i(X,Y)
\left(\sigma^2_{int}(X,Y)\right)_i\right]}{\sum\limits_i
L_i(X,Y)}\right\}^{1/2},
\end{equation}
where $i$ denotes the subsystem and the summation is taken over
all subsystems.
\section{Velocity dispersion profiles of NGC~4594}
In the present section we apply the above constructed  model to a
concrete galaxy. We selected the Sa galaxy NGC~4594 having enough
observational data to construct a detailed mass distribution model.
The velocity dispersions for NGC~4594 have been measured also along
several slit positions outside of the galactic disc.

To avoid calculation errors, we first made several tests: we
calculated dispersions for several simple density distribution
profiles, varied the viewing angle between the disc and the
line-of-sight, varied density distribution parameters. All test
results were in accordance with our physical expectations.

In velocity dispersion calculations all the luminosity distribution
model parameters derived in Section~3 will be handled as fixed. The
visible part of the galaxy is given as a superposition of the
nucleus, the bulge, the disc and the metal-poor halo. The spatial
luminosity and mass density distributions of each visible component
are consistent, i.e. their mass density distribution is given by
\begin{equation}
\rho (a)=\rho (0)\exp [ -( a/(ka_0))^{1/N} ] ,
\end{equation}
where $\rho (0)= l(0)M/L=hM/(4\pi q a_0^3)$ is the central mass
density and $M$ is the component mass. For designations see
Eq.~(\ref{eq1}).

In the case of mass distribution models, a DM component must be
added to visible components. The DM distribution is represented by a
spherical isothermal law
\begin{equation}
\rho(a) = \cases{\rho (0) \{[1+({a\over a_c})^2]^{-1} -
                      [1+({a^0\over a_c})^2]^{-1}\} & $a \leq a^0$
                      \cr
                    0         &      $a>a^0$.    \cr }
\end{equation}
Here $a^0$ is the outer cutoff radius of the isothermal sphere, $a_c
= ka_0$.

Our model includes an additional unknown value -- velocity
ellipsoid orientation. We have to find the best solution to $z_0$,
when fitting the model to the measured dispersions.

\begin{figure}
\includegraphics[width=84mm]{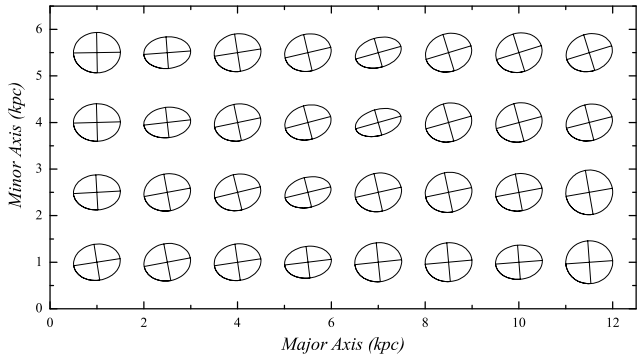}
\caption{Orientation of the calculated velocity dispersion
ellipsoids in galactic meridional plane.}\label{ellipsoid}
\end{figure}

Figure~\ref{ellipsoid} gives the shape and orientation of the
velocity dispersion ellipsoid in the galactic meridional ($R,z$)
plane.

\begin{figure}
\includegraphics[width=84mm]{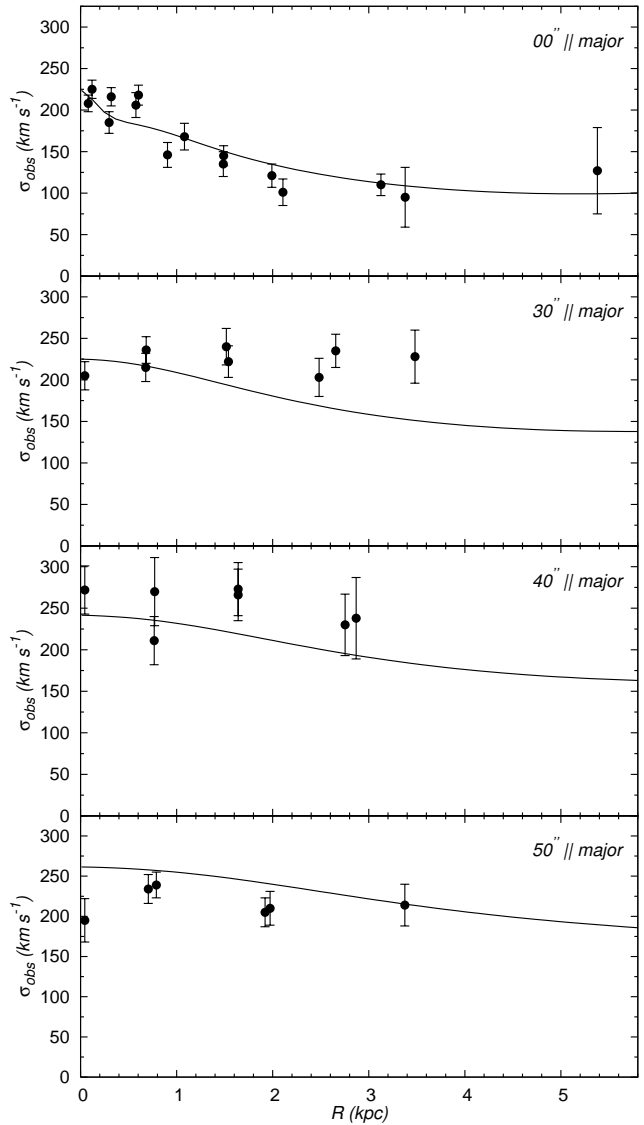}
\caption{Line-of-sight velocity dispersions of NGC~4594 along and
parallel to major axis. Solid line -- calculated model dispersions,
filled circles -- observations.}\label{dispmajor}
\end{figure}

\begin{figure}
\includegraphics[width=84mm]{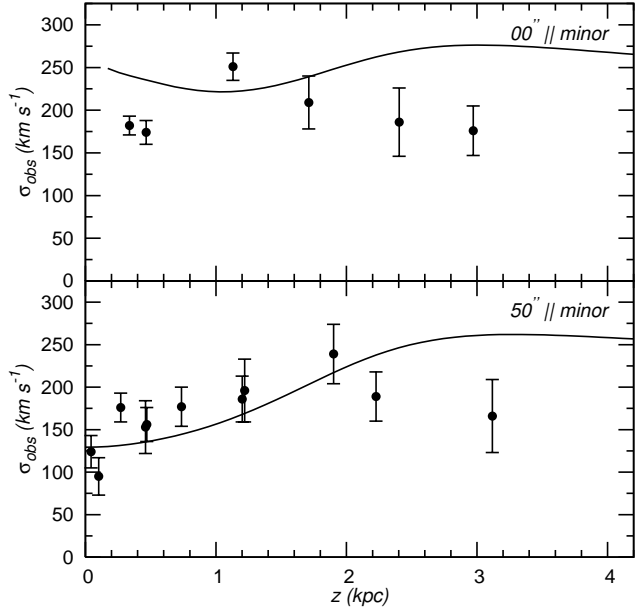}
\caption{Line-of-sight velocity dispersions of NGC~4594 parallel to
the minor axis. Solid line -- calculated model dispersions, filled
circles -- observations.}\label{dispminor}
\end{figure}

In Figure~\ref{dispmajor} calculated line-of-sight dispersions along
and parallel to galactic major axis are given. In
Figure~\ref{dispminor} calculated line-of-sight dispersions parallel
to the minor axis are given.

It is seen that moving further off from the galactic disc, the
results become a little different from the data observed. One reason
may be that we could not find an appropriate solution for $z_0$. It
is possible to fit the data far from the galactic plane with
appropriate selection of $z_0$ but in this case the fit with
dispersions along the major axis is not so good. As a result, the
figures present the best compromise solution we could find.

In addition we must take into account the observed average velocity
dispersion of GCs $\sigma_{\rmn{GC}} = 255~\rmn{km~s^{-1}}$. This
value is clearly higher than the above mentioned last points in
stellar dispersion curve $\sim 160-180~\rmn{km~s^{-1}}$
(Fig.~\ref{dispminor}). All these dispersions correspond to a region
where DM takes effect. For this reason we think that the mean
velocity dispersion of GCs and and stellar velocity dispersions far
outside of galactic plane can be fitted consistently by introducing
a flattened DM halo density distribution. This kind of models have
not yet been constructed by us within the present algorithm.

\begin{figure}
\includegraphics[width=84mm]{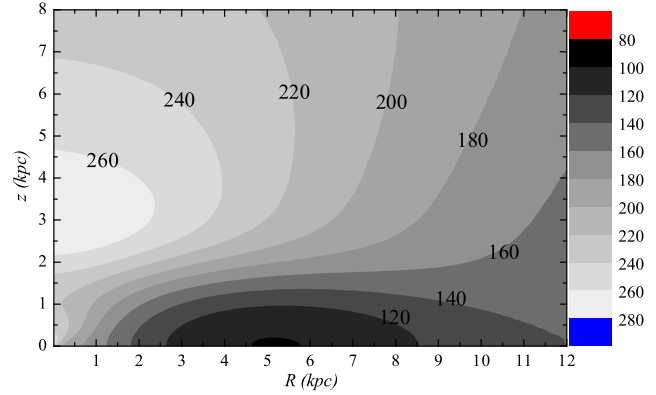}
\caption{Projected line-of-sight dispersions (in $\rmn{km~s^{-1}}$)
in galactic meridional plane.}\label{profile}
\end{figure}

Figure~\ref{profile} gives the calculated velocity dispersion in the
$Rz$-plane illustrating the behavior of dispersions.

\begin{table*}
\caption{Calculated model parameters.} \label{model_param}
\begin{flushleft}
\begin{tabular}{llllllllllll}
\noalign{\smallskip} \hline
Popul. & $M$ & $a_0$ & $q$ & $N$ & $k$ & $h$ & $\beta$ & $L_B$ & $L_V$ & $L_R$ & $L_I$ \\
\noalign{\smallskip} \hline
Nucleus$^a$& 0.001&0.0015&0.99& 3.0 & 0.00297 & 314.3 &     &3.9E-4&      &      & \\
Bulge   & 3.4  & 0.28  & 0.54 & 2.1 & 0.03539 & 44.51 & 0.7 & 0.34 & 0.48 & 0.56 & 0.92 \\
Halo    & 7.4  & 11.0   & 0.75 & 4.0 & 1.465E-4& 2807. & 0.3 & 3.2  &      & 4.4  &    \\
Disc    & 12.0  & 3.4   & 0.25 & 0.78& 0.7429  & 2.607 & 0.88& 1.6  & 2.3  & 2.45 & 5.0  \\
Dark matter& 180  & 40.0  & 1.0   &  & 0.1512  & 14.82 &     &      &      &      & \\
\noalign{\smallskip} \hline
\end{tabular}
\end{flushleft}
\begin{list}{}{}
\item[] Masses and luminosities are in units of $10^{10}\rmn{M_{\sun}}$ and
$10^{10}\rmn{L_{\sun}}$; components radii are in kpc.\\
\item[$^a$] A point mass $10^9\rmn{M_{\sun}}$ have been added to the center of
the galaxy.
\end{list}
\end{table*}

\section{Discussion}

In the present paper we developed an algorithm, allowing to
construct a self-consistent mass and light distribution model and to
calculate projected line-of-sight velocity dispersions outside
galactic plane. We assume velocity dispersion ellipsoids to be
triaxial and thus the phase density is a function of three integrals
of motion. The galactic plane may have an arbitrary angle with
respect to the plane of the sky. The developed algorithm is applied
to construct a mass and light distribution model of the Sa galaxy
M~104. In the first stage a luminosity distribution model was
constructed on the basis of the surface brightness distribution. The
inclination angle of the galaxy is known and the spatial luminosity
distribution can be calculated directly with deprojection. Using the
surface brightness distribution in $BVRI$ colours and along the
major and minor axis, we assume that our components represent real
stellar populations and determine their main structure parameters.
In the second stage, the Jeans equations are solved and the
line-of-sight velocity dispersions and the stellar rotation curve
are calculated. Observations of velocity dispersions outside the
apparent galactic major axis allow to determine the velocity
ellipsoid orientation, anisotropy and to constrain DM halo
parameters.

The total luminosity of the galaxy M~104 resulting from the
best-fitting model is $L_B = (5.1\pm 0.6) \cdot 10^{10}
\rmn{L_{\sun}}$, $L_R = (7.4\pm 0.7) \cdot 10^{10} \rmn{L_{\sun}}$.
The total mass of the visible matter is $M_{\rmn{vis}} = (22.9\pm
3.2) \cdot 10^{10} \rmn{M_{\sun}}$, giving the mean mass-to-light
ratio of the visible matter $M/L_B = 4.5 \pm 1.2~
\rmn{M_{\sun}L_{\sun}^{-1}}$, $M/L_R = 3.1 \pm 0.7~
\rmn{M_{\sun}L_{\sun}^{-1}}$. The surface brightness distributions
in $V$ and $I$ have not sufficient extent to determine the
luminosities of the stellar halo and we do not either give galactic
total luminosities in these colours and corresponding $M/L$ ratios.
Calculated from the model, the $L_B$ coincides well with the total
absolute magnitude $M_B = -21.3$ ($=5.2 \cdot
10^{10}~\rmn{L_\odot}$) obtained by \citet{b39}.

In our model, the mass of the disc is $M_{\rm disc} =
12\cdot10^{10}\rmn{M_{\sun}}$. This coincides rather well with the
disc mass $11.4\cdot 10^{10} \rmn{M_{\sun}}$  calculated with the
help of Toomre's stability criterion by \citet{b83} and with the
mass $9.6\cdot 10^{10} \rmn{M_{\sun}}$ derived by \citet{b34}. On
the other hand, \citet{b34} derived for the bulge mass $> 5\cdot
10^{11} \rmn{M_{\sun}}$, giving $M_\rmn{disc}/M_\rmn{bulge} =0.2$.
This is similar to the value 0.25 derived by \citet{b47}, but this
is much less than $M_\rmn{disc}/M_\rmn{spher} = 1.1$ resulting from
our model. An explanation may be that in the models by \citet{b34}
and \citet{b47} no DM halo was included and hence the extended bulge
mass is higher.

In our model, the disc is rather thick ($q= 0.25$). However, disc
thickness can be easily reduced to $q=0.15-0.2$ when taking the
galactic inclination angle instead of $\delta = 84\degr$ to be
83--82$\degr$. Other parameters remain nearly unchanged. At present
this was not done.

Derived in the present model bulge parameters can be used to compare
them with the results of chemical evolution models. Our model gives
$M/L_V = 7.1 \pm 1.4~\rmn{M_{\sun}L_{\sun}^{-1}}$ and $(B-V)=1.06$
for the bulge. Comparing spectral line intensities with chemical
evolution models, \citet{b87} obtained for the bulge region the
metallicity $Z=0.03$ and the age 11~Gyrs. According to \citet{b13},
these parameters give $M/L_V = 7-8~\rmn{M_{\sun}L_{\sun}^{-1}}$ and
$(B-V) = 1.06-1.08$ for simple stellar population (SSP) models.
Bulge parameters from our dynamical model agree well with these
values and suggest that our model is realistic.

In our calculations, we corrected luminosities from the absorption
in the Milky Way only and did not take into account the inner
absorption in M~104. According to \citet{b35}, absorption in the
centre may be at least $A_V\sim 0.13$~mag and thus $M/L_V = 6.3$ for
the bulge. This is slightly too small when compared with the
\citet{b13} SSP models. However, decreasing the bulge age to
10.5~Gyrs allows to fit the results.

Rather sophisticated models of M~104 have been constructed by
\citet{b34,b35} and \citet{b33}. Due to our different approaches, it
is difficult to compare our components and their parameters with
those of \citet{b34,b35}. On the basis of the data used by us, we
had no reason to add an additional inner disc or a bar to the bulge
region. However, we did not analyse $I$ and $H$ colours and ionized
gas kinematics in inner regions as it was done by \citet{b33}.
Modelling of gas kinematics in central regions is beyond the scope
of the present paper as gas is not collision-free.

On the basis of velocity dispersion observations only along the
major axis it is difficult to decide about the presence of the DM
even when dispersions extend up to 2--3 $\rmn{R_e}$ \citep{b70}. In
the case of M~104, additional dispersion measurements can be used.
Velocity dispersions in the case of the slit positioned parallel and
perpendicular to the galactic major axis, have been measured by
\citet{b53}. The calculated mass distribution model describes rather
well the observed stellar rotation curve and line-of-sight velocity
dispersions. Only the two last measured points at a cut 50~arcsec
perpendicular to the major axis deviate rather significantly when
compared to the model. On the other hand, in addition to stellar
velocity dispersion measurements, the mean line-of-sight velocity
dispersion of the GC subsystem $\sigma = 255~\rmn{km~ s^{-1}}$ was
measured by \citet{b12}. This corresponds to GCs at average
distances 5--10 kpc from the galactic center and is in rather good
agreement with the dispersions calculated from the model.

In the best-fitting model the DM halo harmonic mean radius $a_0=
40$~kpc and $M=1.8\cdot 10^{12} {\rm M_{\sun}}$ giving slightly
falling rotation curve in outer parts of the galaxy
(Fig.~\ref{gasrot}). The central density of the DM halo in our model
is $\rho (0) = 0.033~\rmn{M_{\sun} pc^{-3}}$, being also slightly
less than it was derived for distant ($z\sim ~0.9$) galaxies
\citep[$\rho (0) = 0.012-0.028~\rmn{M_{\sun} pc^{-3}}$,][]{b78}. On
the other hand, the result fits with the limits derived by
\citet{b9} for local galaxies $\rho (0) = 0.015-0.050~\rmn{M_{\sun}
pc^{-3}}$.

An essential parameter in mass distribution determination is the
inclination of the velocity dispersion ellipsoid with respect to the
galactic plane \citep[see e.g.][]{b56,b63}. Velocity dispersion
ellipsoid inclinations calculated in the present paper are moderate,
being $\le 30\degr$. In a sense, our approach to the third integral
of stellar motion is similar to that by \citet{b50} -- the local
St\"ackel fit. In their modelling of the local Milky Way structure,
they derived that at $0<z<600$ pc and $6.8<R<8.8$ kpc, the
inclination of the velocity dispersion ellipsoid is less than $z/R$,
and they studied the corresponding correction in detail. In our
model, in the same distance regions (although the Milky Way and
M~104 are not very similar objects), inclination correction values
are slightly smaller. Variations of the corrections with $R$ and $z$
are qualitatively similar. The largest difference is the variation
of the correction value with $z$, for which \citet{b50} obtained an
increase by 0.1, when moving from $z=0$ to $z= 0.6$~kpc, but in our
model, corresponding increase was only by 0.01. In our model, a
significant increase of the ellipsoid inclination angle begins at
larger $z$, which can be explained by higher thickness of the disc
component of M~104 ($q=0.25$). In a following paper we intend to
construct similar models for other galaxies with velocity dispersion
measurements outside the galactic major axis. This may lead to more
firm conclusions about the inclination of velocity dispersion
ellipsoids outside the galactic plane.

\begin{figure}
\includegraphics[width=84mm]{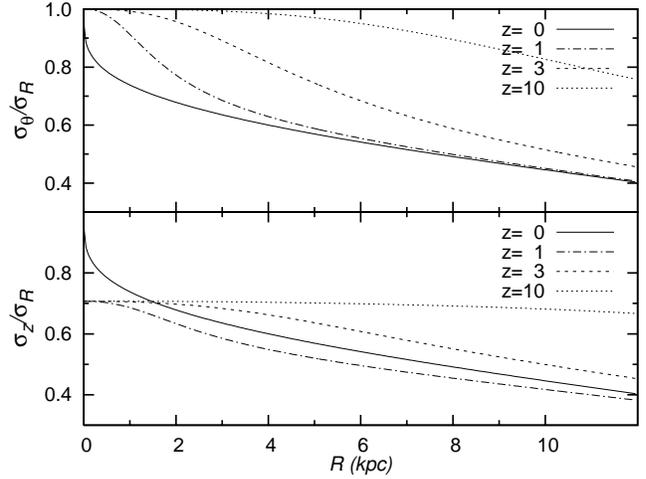}
\caption{Dispersion ratios in galactic meridional plane. Coordinate
$z$ is in kpc.}\label{aniso}
\end{figure}

Ratios of the line-of-sight velocity dispersions are given in
Fig.~\ref{aniso}. It is seen that velocity dispersion ellipsoids are
quite elongated -- anisotropies in the symmetry plane at outer parts
of the galaxy are $<0.5$.

Modelling the disc Sb galaxy NGC~288 within a constant velocity
ellipsoid inclination approximation, \citet*{b43} estimated that the
dispersion ratio $\sigma_z/\sigma_R = 0.70$. Within epicycle
approximation \citet{b89} derived for Sb galaxy NGC~3982 the
dispersion anisotropy $\sigma_z/ \sigma_R = 0.73$. These two
galaxies are morphologically close to the Sa galaxy modelled in the
present paper, and it is seen that dispersion ratios are more
anisotropic in our case. In general, a radially elongated dispersion
ellipsoid is rather common \citep*{b75}. On the other hand, there
are exceptions -- galaxy NGC~3949 (\citet{b89} has the dispersion
ratio $\sigma_z/\sigma_R = 1.18$).

By using the quadratic programming method \citep{b23}, the
distribution function within the three-integral approximation has
been numerically calculated for the S0 galaxy NGC~3115 by
\citet*{b36}. Assuming some similarity between S0 and Sa galaxies,
it is interesting to compare the derived velocity dispersion
behaviour outside the galactic plane. Although detailed comparison
is difficult a similar structure of isocurves is seen. But the
results disagree in values of $\sigma_z/ \sigma_R$. Our ratios are
radially elongated, the ratios by \citet{b36} are vertically
elongated. Qualitatively this is in agreement with the general trend
that galaxies of earlier morphological type have larger $\sigma_z/
\sigma_R$ ratio \citep{b75}. At larger $R$, the dispersion ratio
$\sigma_z/ \sigma_R$ decreases. This is in agreement with the
decrease of dispersion ratios due to the decrease of the role of
interactions with molecular clouds at greater galactocentric
distances \citep[see][]{b48}. The use of this explanation in the
case of gas-poor S0 galaxies is not clear. At greater distances $z$
both our and \citet{b36} dispersion ellipsoids become more
spherical.

By using the Schwarzschild method, dispersion ratios for E5--6
galaxy NGC~3377 have been calculated by \citet*{b20}. Taking into
account relation between spherical and cylindrical coordinates, the
behaviour of the dispersion ratios as a function of $R$ and $z$ near
the galactic plane is in approximate accordance (dispersion ratios
by \citet{b20} are slightly more spherical). Rotational properties
of elliptical and Sa galaxies are too different to compare the
ratios $\sigma_{\theta}/ \sigma_R$. Unfortunately, it is not
possible to compare also orientations of velocity dispersion
ellipsoids.

\section*{Acknowledgements}
We thank Dr. U. Haud for making available his programs for the light
distribution model calculations. We would like to thank the
anonymous referee for useful comments and suggestions helping to
improve the paper. We acknowledge the financial support from the
Estonian Science Foundation (grants 4702 and 6106). This research
has made use of the NASA/IPAC extragalactic database (NED), which is
operated by the Jet Propulsion Laboratory, California Institute of
Technology, under contract with the NASA.

\label{lastpage}

\end{document}